\documentclass[12pt]{article}
\pdfoutput=1
\usepackage{subfigure}
\usepackage{amssymb,amsmath}
\usepackage{graphicx}
\usepackage{color}
\usepackage[colorlinks=true
,urlcolor=blue
,citecolor=blue
,linkcolor=blue
,pagecolor=blue
,linktocpage=true
,pdfproducer=medialab
]{hyperref}
\usepackage[a4paper,width=15.2cm]{geometry}
\makeatletter \renewcommand{\@dotsep}{10000} \makeatother

\newcommand{\beq}{\begin{equation}}
\newcommand{\eeq}{\end{equation}}
\newcommand{\bea}{\begin{eqnarray}}
\newcommand{\eea}{\end{eqnarray}}

\begin{document}
%Remove date before submitting to arXiv
%\date{\today}

\begin{center}

 {\Large\bf Sparticle Spectroscopy with Neutralino Dark matter from  $t$-$b$-$\tau$ Quasi-Yukawa Unification
 } \vspace{1cm}

{\large  Shahida Dar$^{a,}$\footnote{ E-mail:
shahida.dar@mvcc.edu}, Ilia Gogoladze$^{b,}$\footnote{E-mail:
ilia@bartol.udel.edu\\ \hspace*{0.5cm} On  leave of absence from:
Andronikashvili Institute of Physics, 0177 Tbilisi, Georgia.}, Qaisar Shafi$^{b}$ and
Cem Salih $\ddot{\rm U}$n $^{b,}$\footnote{ E-mail: cemsalihun@bartol.udel.edu}} \vspace{.5cm}

{\baselineskip 20pt \it
$^a$Mohawk Valley Community College, 1101 Sherman Dr. Utica, NY 13501, USA \\
$^b$Bartol Research Institute, Department of Physics and Astronomy, \\
University of Delaware, Newark, DE 19716, USA  } \vspace{.5cm}

\vspace{1.5cm}
 {\bf Abstract}
\end{center}

We consider two classes of $t$-$b$-$\tau$ quasi-Yukawa unification scenarios which can arise from realistic supersymmetric SO(10) and $ SU(4)_{C}\times SU(2)_{L}\times SU(2)_{R} $ models. We show that these scenarios can be successfully implemented in the CMSSM and NUHM1 frameworks, and yields a variety of sparticle spectra with WMAP compatible neutralino dark matter. In NUHM1 we find bino-higgsino dark matter as well as the stau coannihilation and $A$-funnel solutions. The CMSSM case yields the stau coannihilation and $A$-funnel solutions. The gluino and squark masses are found to lie in the TeV range.

\newpage

%%%%%%%%%%%%%%%%%%%%%%%%%%%%%%%%%%%%%%%%%%%%%%%%%%%%%%%%%%%%
\renewcommand{\thefootnote}{\arabic{footnote}}
\setcounter{footnote}{0}

%%%%%%%%%%%%%%%%%%%%%%%%%%%%%%%%%%%%%%%%%%%%%%%%%%%%%%%%%%%%%

%\baselineskip 36pt
% Main body
%%%%%%%%%%%%%%%%%%%%%%%%%%
%\baselineskip 18pt
%%%%%%%%%%%%%%%%%%%%%%%%%%

\section{\label{ch:introduction}Introduction}

Third family ($t$-$b$-$\tau$) Yukawa Unification (YU, for short) at the GUT scale
 $ M_{\rm G}~( \sim 3\times 10^{16} {\rm GeV}) $
 is predicted by the simplest supersymmetric SO(10) GUT if the MSSM Higgs doublets are assumed to reside in the Higgs 10-plet \cite{big-422}. The implications of YU for Higgs and sparticle spectroscopy have been extensively considered in the literature \cite{bigger-422}.
 More recently \cite{Baer:2008jn},
 % \cite{Gogoladze:2009bn, Gogoladze:2011db},
it has been argued  that SO(10) GUT YU  predicts
 relatively light ($m_{\tilde{g}}<$ TeV) gluinos, which can
be readily tested \cite{Baer:2009ff}  at the Large Hadron Collider (LHC). The squarks and sleptons
turn out to have masses in the multi-TeV range.

In order to reconcile radiative electroweak symmetry breaking  (REWSB) with YU, the MSSM Higgs
 soft supersymmetry breaking (SSB) masses must be split in
such way  that $m^2_{H_{d}}/ m^2_{H_u}> 1.2$  at  $M_{\rm G}$ \cite{Olechowski:1994gm}.
As mentioned above, the MSSM doublets reside in the 10 dimensional
representation of SO(10) GUT  for YU
condition to hold. In the gravity mediated
 supersymmetry breaking  scenario    \cite{Chamseddine:1982jx} the required splitting in the Higgs sector can be generated by involving additional Higgs fields \cite{Gogoladze:2011db},
  or via D-term contributions  \cite{Drees:1986vd}.
 Note that  YU  is sensitive not only to the value of $\tan \beta$, but also to weak
 scale threshold corrections   \cite{Gogoladze:2011db, Gogoladze:2009bn}.

  On the other hand, one  knows that  a singlet 10-plet of Higgs field   does not work for the first
  two generations of quarks and leptons.
   One way to fix this problem in SO(10)  is to  extend the Higgs sector
   which couples to the SM fermions, in particular by introducing  Higgs 126-plet \cite{Lazarides:1980nt}.
    In this case,  the low energy MSSM
   Higgs doublets are a linear superposition of various SO(10) Higgs fields. Depending on the parameters, this may lead to deviation from exact $t$-$b$-$\tau$ YU.  As pointed out in ref. \cite{Gomez:2002tj}, a relatively small  deviation from  $t$-$b$-$\tau$ YU (referred to here as quasi-YU)
 allows REWSB with universal SSB mass terms for the MSSM Higgs fields at $M_{\rm G}$.
  In this paper we revisit and expand the analysis presented in \cite{Gomez:2002tj}.
We find that a modest relaxation of    $t$-$b$-$\tau$ YU  condition within the SO(10) GUT framework allows us to
significantly lower sfermion masses which can be tested at the LHC. The quasi-YU (QYU) framework also allows one to implement the neutralino dark matter scenario consistent with the  Wilkinson Microwave Anisotropy Probe (WMAP)  \cite{Komatsu:2008hk} constraints. This is not possible, it appears, in SO(10) models with exact YU \cite{Baer:2008jn}.

This paper is organized as follows.
In Section \ref{model}, we present an SO(10) model which naturally yields QYU.
In Section \ref{constraintsSection} we describe the
scanning procedure and various SUSY constraints
imposed on the parameter space of NUHM1 (non-universal Higgs model with $ m_{H_{u}}=m_{H_{d}}\neq m_{0} $) and CMSSM (constrained minimal sumersymmetric model).
In Sections \ref{ch:results} and \ref{ch:cmssm} and we present our results and highlight some benchmark points
of QYU condition. The correlation between direct and
indirect detection of dark matter and the QYU condition is presented in Section \ref{ch:dark}.
 Our conclusions are summarized in
Section \ref{ch:conclusions}.

\section{The Model \label{model}}

One  way to obtain the correct fermion masses and mixings in SO(10) GUT
 is to utilize Higgs  in the 10 and 126 dimensional representations. The Yukawa interactions in this case are given by
 \begin{equation}
Y^{i j}_{10}~16_i ~.~ 16_j  ~.~  H_{10} +
Y^{i j}_{126}~16_i ~.~  16_j ~.~ H_{\overline{126}} \,,  \label{eq1}
 \end{equation}
where $Y^{i j}_{10}$ and $Y^{i j}_{126} $ denote the Yukawa couplings. From the coupling between the
$\overline{126}$ and the {10}-plet Higgs,  the SM doublet
fields contained in $\overline{126}$ will acquire vacuum
expectation values (VEVs) through mixing with the VEVs of the Higgs doublets
in ${10}$ \cite{Lazarides:1980nt}.
 The modification  \cite{Gomez:2002tj}  of  $t$-$b$-$\tau$ YU  condition
  depends on how the doublets from the $\overline{126}$ and  $10$
 mix and the values of Yukawa couplings in Eq.  (\ref{eq1}).
One possible mixing of these doublets arises from the following interaction
    \begin{equation}
 \lambda_1  ~.~ 10 ~.~ \overline{126}  ~.~  210 + \lambda_2  ~.~ 10  ~.~ {126}
 ~.~  210 , \label{210h}
  \end{equation}
   where $\lambda_1$ and $\lambda_2$ are dimensionless couplings.
  The 210-plet has  an $M_{\rm G}$ scale VEV and is primarily used for breaking SO(10) to its maximal subgroup $ SU(4)_{c}\times SU(2)_{L}\times SU(2)_{R} $ \cite{Pati:1974yy}.
  However, we will exploit here the fact that there exist other $M_{\rm G}$ scale VEV directions in the 210-plet.
  Let us
   decompose  the interaction in Eq. (\ref{210h}) in
   terms of the $SU(4)_c\times SU(2)_L \times SU(2)_R$ symmetry \cite{Bajc:2004xe}
\begin{align}
   \lambda_1  (1,2,2)_{10} (15,2,2)_{\overline{126}} [(15,1,3)_{210}+ (15,1,1)_{210}]  \nonumber \\ \nonumber \\
    +  \lambda_2 (1,2,2)_{10}(15,2,2)_{{126}} [(15,1,3)_{210}+ (15,1,1)_{210}]\label{210dec} + \ldots
\end{align}
 Here  we list only the relevant couplings and for simplicity,
 we do not consider the mixing of Higgs doublets  from the 210  and 10 which does not provide any contribution to the Yukawa sector.

 It was pointed out in \cite{Bajc:2004xe}  that it  is possible to  develop a VEV in the directions
  $(15,1,3)_{210}$ or  $(15,1,1)_{210}$,  or simultaneously in both directions.
   We assume that these VEVs are order of $M_{\rm G}$.
   After fine tuning, one pair of Higgs doublets can be identified as the
MSSM pair ($H_u+H_d$), which, as previously stated, is an admixture of  Higgs
doublets from the 10 and 126. The other scalar doublets have masses
of order $M_{\rm G}$. With a non-zero VEV along the $(15,1,3)_{210}$  direction, the coupling
  $(1,2,2)_{10} (15,2,2)_{\overline{126}} (15,1,3)_{210}$ generates $SU(2)_{R}$
  violating bilinear terms between the up and down type Higgs doublets, which
  effectively violates top-bottom  YU condition at $M_{\rm G}$.  In this case, following closely the $ SU(4)_{C}\times SU(2)_{L}\times SU(2)_{R} $ discussion in \cite{Gomez:2002tj}, one can derive  the following asymptotic relations among the three Yukawa couplings:
\begin{align}
y_t:y_b:y_{\tau}=(1+C):(1-C):(1+3C) \label{rela1}  ,\rm
~~~~~~~~~~~~~{(Case ~~I)}
   \end{align}
where $C$ is taken to be real and positive. In practice, we will find that $ C \sim 0.1-0.2 $, and we
   refer to the QYU condition in Eq. (\ref{rela1}) as Case I. Note that REWSB is easier to achieve for $ C > 0 $.

For a slightly different scenario, consider the case in which a
VEV is developed  only  in the  $(15,1,1)_{210}$  direction in Eq. (\ref{210dec}).
The mixing of Higgs  doublets from   the interaction   $(1,2,2)_{10} (15,2,2)_{\overline{126}} (15,1,1)_{210}$
   is $ SU(2)_R$ invariant, and  at $M_{\rm G}$, the top-bottom YU
  condition still holds. In this case one finds the relation
  \begin{align}\hspace{-3.2cm}
y_t:y_b:y_{\tau}=(1+C^{\prime}):(1+C^{\prime}):(1-3C^{\prime}), \label{rela1a}
   \end{align}
where $C^{\prime}$ has the same definition as $C$, but numerically it can be different.

  If the VEVs develop along both the $(15,1,1)_{210}$
   and $(15,1,3)_{210}$  directions, we simply add  Eqs.  (\ref{rela1}) and (\ref{rela1a}) to get \cite{Gomez:2002tj}
  \begin{align}\hspace{0.4cm}
y_t:y_b:y_{\tau}=(1+C_1):(1-C_2):(1+3C_2), \label{rela2} \rm
~~~~~~~~~~~~~{(Case ~~II)}
   \end{align}
  where $ C_{1}=C+C'$, $ C_{2}=C-C'$. The QYU relation given in Eq. (\ref{rela2}) will be referred to as Case II.

\section{Phenomenological constraints and scanning procedure\label{constraintsSection}}

We employ the ISAJET~7.80 package~\cite{ISAJET}  to perform random
scans over the fundamental parameter space. In this package, the weak scale values of gauge and third generation Yukawa
couplings are evolved to $M_{\rm G}$ via the MSSM renormalization
group equations (RGEs) in the $\overline{DR}$ regularization scheme.
We do not strictly enforce the unification condition $g_3=g_1=g_2$ at $M_{\rm
G}$, since a few percent deviation from unification can be
assigned to unknown GUT-scale threshold
corrections~\cite{Hisano:1992jj}.
The deviation between $g_1=g_2$ and $g_3$ at $M_{G}$ is no
worse than $3-4\%$.
For simplicity  we do not include the Dirac neutrino Yukawa coupling
in the RGEs, which is expected to be small.

The various boundary conditions are imposed at
$M_{\rm G}$ and all the SSB
parameters, along with the gauge and Yukawa couplings, are evolved
back to the weak scale $M_{\rm Z}$.
In the evaluation of Yukawa couplings the SUSY threshold
corrections~\cite{Pierce:1996zz} are taken into account at the
common scale $M_{\rm SUSY}= \sqrt{m_{{\tilde t}_L}m_{{\tilde t}_R}}$. The entire
parameter set is iteratively run between $M_{\rm Z}$ and $M_{\rm
GUT}$ using the full 2-loop RGEs until a stable solution is
obtained. To better account for leading-log corrections, one-loop
step-beta functions are adopted for gauge and Yukawa couplings, and
the SSB parameters $m_i$ are extracted from RGEs at multiple scales
$m_i=m_i(m_i)$. The RGE-improved 1-loop effective potential is
minimized at an optimized scale $M_{\rm SUSY}$, which effectively
accounts for the leading 2-loop corrections. Full 1-loop radiative
corrections are incorporated for all sparticle masses.

The requirement of REWSB  puts an important theoretical
constraint on the parameter space. Another important constraint
comes from limits on the cosmological abundance of stable charged
particles  \cite{Nakamura:2010zzi}. This excludes regions in the parameter space
where charged SUSY particles, such as ${\tilde \tau}_1$ or ${\tilde t}_1$, become
the lightest supersymmetric particle (LSP). We accept only those
solutions for which one of the neutralinos is the LSP and saturates
the WMAP  dark matter relic abundance bound.

We have performed random scans for the following parameter range:
\begin{align}
0\leq  m_{0}  \leq 20\, \rm{TeV} \nonumber \\
0\leq   m_{H_u}= m_{H_d} \leq 20\, \rm{TeV} \nonumber \\
0 \leq M_{1/2}  \leq 3  \rm{TeV} \nonumber \\
45\leq \tan\beta \leq 60 \nonumber \\
-3\leq A_{0}/m_0 \leq 3
 \label{parameterRange}
\end{align}
with  $\mu > 0$ and  $m_t = 173.1\, {\rm GeV}$  \cite{:1900yx} with $ m_{0}\neq m_{H_{u},H_{d}} $, this is usually referred to as NUHM1 \cite{Baer:2008ih}. This choice of parameter space  was informed by our previous experience studying
exact  $t$-$b$-$\tau$ YU \cite{ Gogoladze:2011db,Gogoladze:2009bn}. In section \ref{ch:cmssm}, we will consider the well-known case of CMSSM with $ m_{0}=m_{H_{u},H_{d}} $. In contrast to NUHM1, we are unable to identify  bino-Higgsino dark matter in the CMSSM framework with QYU.

Our results are not
too sensitive to one or two sigma variation in the value of $m_t$  \cite{Gogoladze:2011db}.
We use $m_b(m_Z)=2.83$ GeV which is hard-coded into ISAJET.

In scanning the parameter space, we employ the Metropolis-Hastings
algorithm as described in \cite{Belanger:2009ti}. All of the
collected data points satisfy
the requirement of REWSB,
with the neutralino in each case being the LSP. Furthermore,
all of these points satisfy the constraint $\Omega_{\rm CDM}h^2 \le 10$.
This is done so as to collect more points with a WMAP compatible value of cold dark
matter (CDM) relic abundance. For the Metropolis-Hastings algorithm, we only use
the value of $\Omega_{\rm CDM}h^2$ to bias our search. Our purpose in using the
Metropolis-Hastings algorithm is to be able to search around regions of
acceptable $\Omega_{\rm CDM}h^2$ more fully. After collecting the data, we impose
the mass bounds on all the particles \cite{Nakamura:2010zzi} and use the
IsaTools package~\cite{Baer:2002fv}
to implement the following phenomenological constraints: We apply the following experimental constraints successively on the data that
we acquire from ISAJET:
\begin{table}[h]\centering
\begin{tabular}{rlc}
$m_h~{\rm (lightest~Higgs~mass)} $&$ \geq\, 114.4~{\rm GeV}$          &  \cite{Schael:2006cr} \\
$BR(B_s \rightarrow \mu^+ \mu^-) $&$ <\, 5.8 \times 10^{-8}$        &   \cite{:2007kv}      \\
$2.85 \times 10^{-4} \leq BR(b \rightarrow s \gamma) $&$ \leq\, 4.24 \times 10^{-4} \;
 (2\sigma)$ &   \cite{Barberio:2008fa}  \\
$0.15 \leq \frac{BR(B_u\rightarrow
\tau \nu_{\tau})_{\rm MSSM}}{BR(B_u\rightarrow \tau \nu_{\tau})_{\rm SM}}$&$ \leq\, 2.41 \;
(3\sigma)$ &   \cite{Barberio:2008fa}  \\
$\Omega_{\rm CDM}h^2 $&$ =\, 0.111^{+0.028}_{-0.037} \;(5\sigma)$ &
\cite{Komatsu:2008hk}
\end{tabular}
\end{table}

For $BR(B_{s}\rightarrow s\gamma)$ we use 2$\sigma$ significance because it is relatively well measured
by several experiments. Moreover, it is commonly used by several authors, and so we have considered the constraint on $BR(B_{s}\rightarrow s\gamma)$ to the same significance order as some other studies.

We use 3$\sigma$ significance for $\frac{BR(B_u\rightarrow \tau \nu_{\tau})_{\rm MSSM}}{BR(B_u\rightarrow \tau \nu_{\tau})_{\rm SM}}$
because this ratio suffers from large uncertainties in the determination of $\lvert V_{ub} \rvert$ . It is also commonly used in the literature.

We use 5$\sigma$ significance for $\Omega_{\rm CDM}h^2 $ constraints because the co-annihilation process for calculating the relic abudance is
an exponential function of the difference in masses between the LSP and the NLSP. A slight change in this mass difference can produce large uncertainties. We therefore think that 5$\sigma$ is an
appropriate and conservative range to take for $\Omega_{\rm CDM}h^2 $.

%%%%%%%%%%%%%%%%%%%%%%%%%%%%%%%%%%%%%%%%%%%%%%%%%%%%%%%%%%%%%%%%%%%%%%%%%%%%%%

\section{Quasi-Yukawa Unification and Sparticle Spectroscopy in NUHM1 \label{ch:results}}
\subsection{Case~~I}

\begin{figure}[t!]
\centering
\includegraphics[width=15cm]{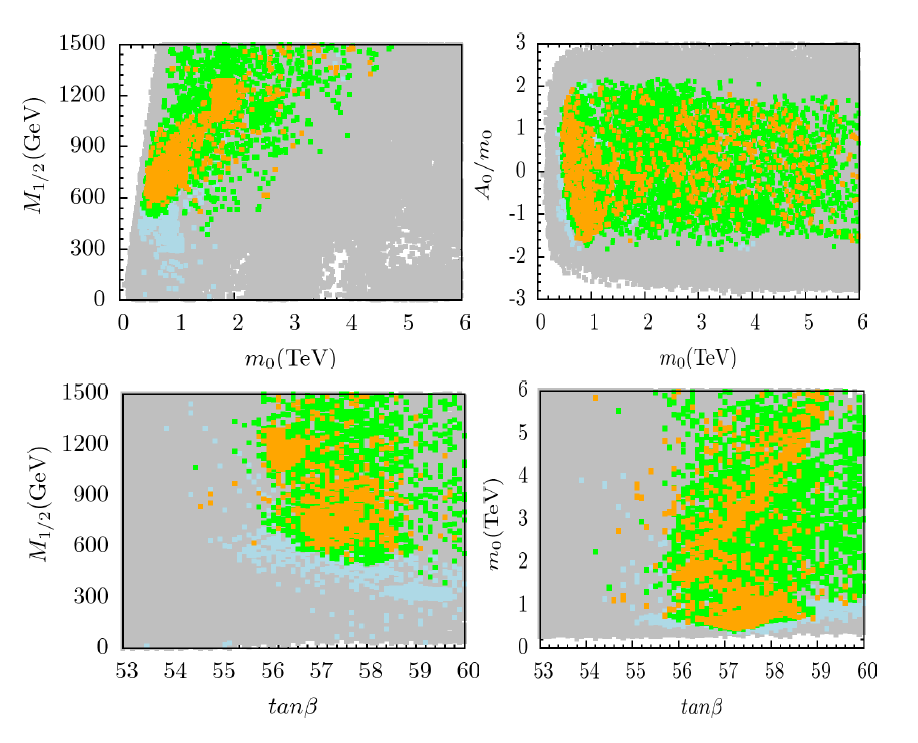}
\caption{ Plots in the $M_{1/2}-m_0$, $M_{1/2}- \tan\beta$,
$A_0/m_0-m_0$, $m_0-\tan\beta$  planes for Case I. Gray points are consistent
with REWSB and $\chi^0_1$ LSP. Light blue points satisfy the QYU constraint given in Eq. (\ref{rela1}). The green
points  satisfy  particle mass bounds and constraints from
$BR(B_s\rightarrow \mu^+ \mu^-)$, $BR(B_u\rightarrow \tau
\nu_{\tau})$ and $BR(b\rightarrow s \gamma)$. In addition, we
require that these points do no worse than the SM in terms of the
$(g-2)_\mu$ prediction. Yellow points belong to the subset of green
points that satisfies all constraints including dark matter ones from WMAP.
\label{plots1}}
\end{figure}

In Fig.~\ref{plots1} we present our results in the
$M_{1/2}-m_0$, $M_{1/2}-\tan\beta$, $A_0/m_0-m_0$,
$m_0-\tan\beta$  planes. The gray points are consistent with REWSB and
$\chi^0_1$ LSP, and the light blue points satisfy the QYU constraint given in
 Eq. (\ref{rela1}). The green  points are a subset of blue points and satisfy  particle mass bounds and
constraints from $BR(B_s\rightarrow \mu^+ \mu^-)$,
$BR(B_u\rightarrow \tau \nu_{\tau})$ and $BR(b\rightarrow s
\gamma)$. In addition, we require that these points do no worse than
the SM in terms of the $(g-2)_\mu$ prediction. The yellow points belong
to the
 subset of green points that satisfy all constraints including the WMAP observed dark matter density.
From the $M_{1/2}-\tan\beta$ and  $m_0-\tan\beta$
planes we see that realistic solutions arise for
 $m_0 \gtrsim 500$ GeV and $M_{1/2} \gtrsim 600$ GeV.
The $A_0/m_0-m_0$ plane shows that in contrast to $t$-$b$-$\tau$
YU, QYU does not have a preferred value
for $A_0/m_{0}$ \cite{Baer:2008jn}, and viable solutions can be obtained for
$ \left| A_{0}/m_{0}\right| \gtrsim 2  $.

\begin{figure}[t!]
\centering
\includegraphics[width=15cm]{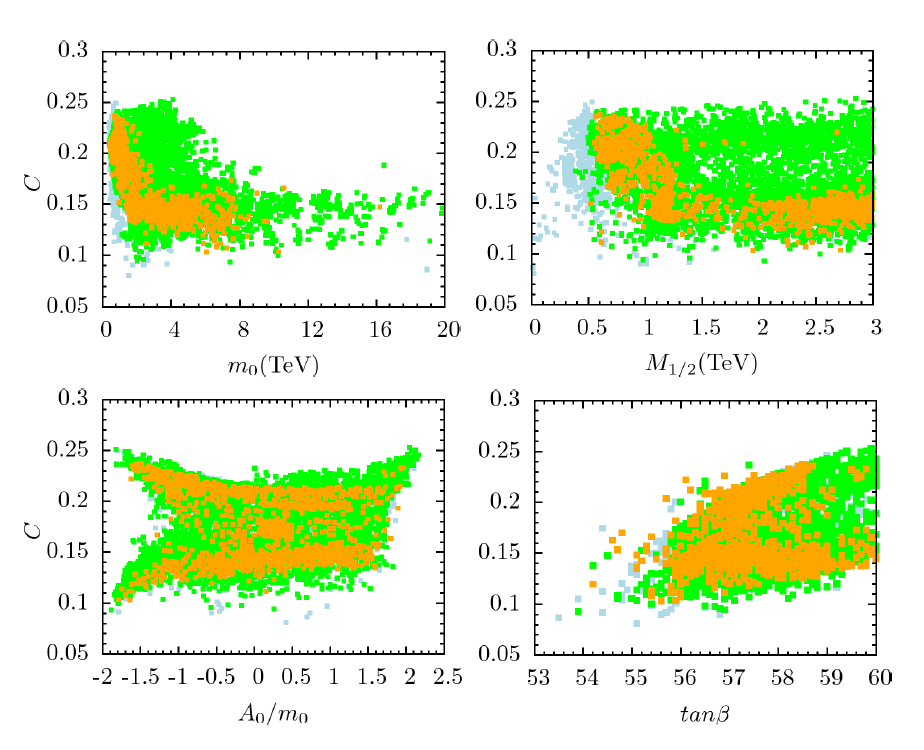}
\caption{
Plots in the $C-m_0$, $C-M_{1/2}$, $C-A_0/m_0$, $C-\tan\beta$  planes for Case I.
Color coding is the same as in Fig. \ref{plots1}.
\label{plots2}}
\end{figure}

In Fig. \ref{plots2}, we present some results pertaining to the parameter $ C $ with the same color coding as Fig.~\ref{plots1}.  We observe that $C$ as small as
0.12 is compatible with all experimental constraints (yellow points). The
lower bound on $C$ is dictated mostly by the REWSB condition. As
mentioned above, we have a universal Higgs SSB bilinear term
($m_{H_u}= m_{H_d}$) at $M_{\rm G}$. In this case, for
REWSB the Yukawa coupling $y_t$ has to be larger than the Yukawa coupling $y_b$
between $M_{\rm G}$  and $M_Z$. In the data that we have collected satisfying the QYU condition
in the NUHM1 parameter space, we find that $y_t-y_b \gtrsim 0.1$. This, according to
Eq. (\ref{rela1}), is equivalent to $C \gtrsim 0.12$.

It was pointed out in ref. \cite{Gogoladze:2011db} that it is  hard to have
$y_t>0.6$. To see this let us consider the SUSY
threshold corrections to the top quark mass.
 The leading correction is given by \cite{Pierce:1996zz}
\begin{align}
\delta y_t^{\rm finite}\approx\frac{g_3^2}{12\pi^2}\frac{\mu
m_{\tilde g} \tan\beta}{m_{\tilde t}^2}~. \label{t-thres}
\end{align}
In our sign convention (evolving the couplings from $M_{\rm G}$ to
$M_Z$), a negative contribution to $\delta y_t$ is preferred.
Naively, a larger negative contribution allows for a larger $y_{t}
(M_{\rm G})$. However, in the case of same sign gauginos with
$\mu>0$, we get a positive contribution to $\delta y_t$, thus a large $m_0$ value is required in order to minimize the
contribution coming from Eq. (\ref{t-thres}).
 The significance of looking at the sign
of the correction to $\delta y_t$ in this case is the realization
that it may not be possible to achieve $y_t>0.6$.
 On the other hand, despite the possibility for large thresholds for the large bottom quark, one may not have
arbitrarily small values for $ y_{b} $ at $M_{\rm G}$. From the
data we find that $y_b(M_{\rm G})$ cannot be much
smaller then 0.35 or so.  If we use the maximal value for  $y_t$  and the minimal value for $y_b$
in the expression  $C=(y_t-y_b)/(y_t+y_b)$ (which can be derived from   Eq. (\ref{rela1})),  we see that $C \gtrsim 0.25$.

\begin{figure}[t!]
\centering
\includegraphics[width=15cm]{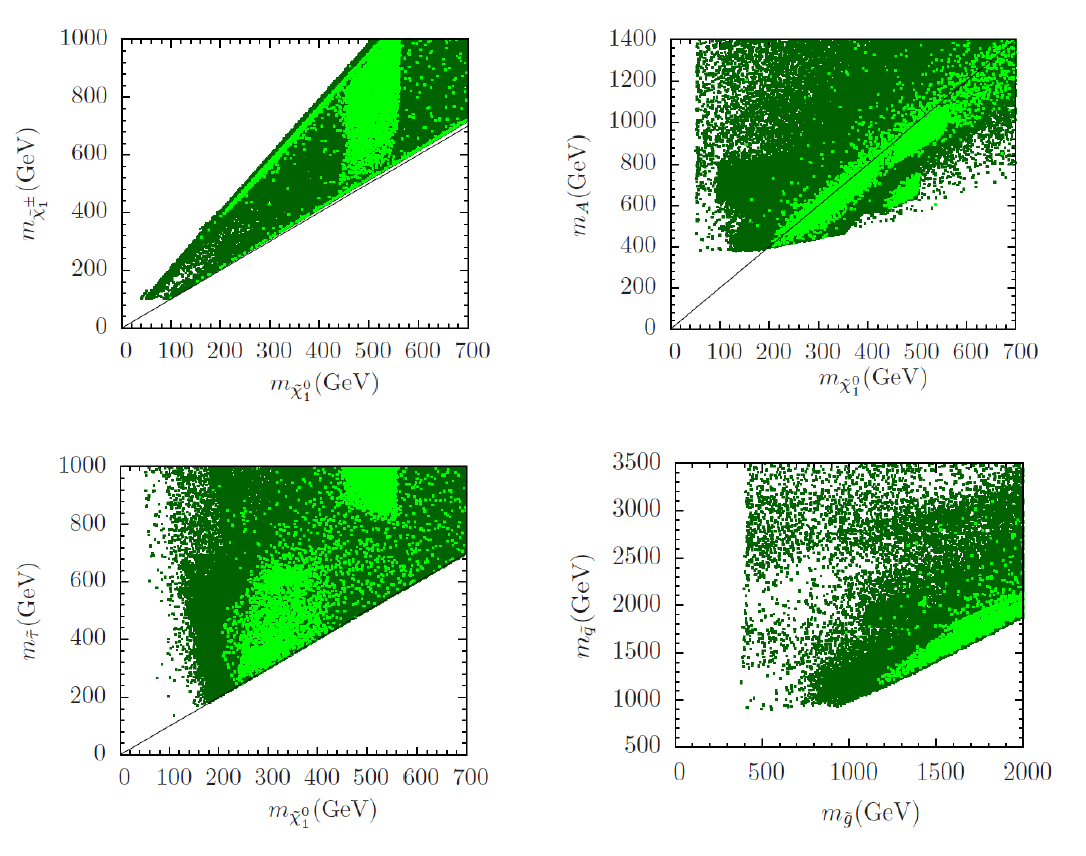}
\caption{
Plots in the $m_{\tilde{\chi}_1^{\pm}} - m_{\tilde{
\chi}_1^0}$, $m_{A} - m_{\tilde{ \chi}_1^0}$,
$m_{\tilde{\tau}} - m_{\tilde{ \chi}_1^0}$ and
$m_{\tilde {q}} - m_{\tilde{ g}}$  planes for Case I.
The dark green points in this figure satisfy the
requirements of REWSB,  $\tilde{\chi}^0_{1}$ LSP,  the particle mass bounds and constraints from
$BR(B_s\rightarrow \mu^+ \mu^-)$, $BR(B_u\rightarrow \tau
\nu_{\tau})$ and $BR(b\rightarrow s \gamma)$  in NUHM1. In addition, we
require that these points do no worse than the SM in terms of the
$(g-2)_\mu$ prediction. The light green  points
are consistent with QYU condition and constraints mentioned above.
We show in the $m_{\tilde{\chi}_1^{\pm}} - m_{\tilde{\chi}_1^0}$,  and $m_{\tilde
{\tau}} - m_{\tilde{ \chi}_1^0}$ planes the unit slope lines that
indicate the respective coannihilation channels. In the
$m_A - m_{\tilde{\chi}_1^0}$ plane we show the line
$m_A=2m_{\tilde{\chi}_1^0}$ that signifies the
$A$-funnel solution.
\label{plots3}}
\end{figure}

In Fig.~\ref{plots3}, we show the relic density channels consistent
with QYU condition (Eq. (\ref{rela1})) in the $m_{\tilde{\chi}_1^{\pm}} - m_{\tilde{
\chi}_1^0}$,  $m_{\tilde
{\tau}} - m_{\tilde{ \chi}_1^0}$ and $m_A - m_{\tilde{
\chi}_1^0}$ planes. The dark green points in this figure satisfy the
requirements of REWSB,  $\tilde{\chi}^0_{1}$ LSP,  the particle mass bounds and constraints from
$BR(B_s\rightarrow \mu^+ \mu^-)$, $BR(B_u\rightarrow \tau
\nu_{\tau})$ and $BR(b\rightarrow s \gamma)$. In addition, we
require that these points do no worse than the SM in terms of the
$(g-2)_\mu$ prediction. The light green  points
are consistent with QYU condition and the constraints mentioned above.
We can see in
Fig.~\ref{plots3} that a variety of coannihilation and
annihilation scenarios are compatible with QYU and
neutralino dark matter. Included in the $m_A - m_{\tilde{
\chi}_1^0}$ plane is the line $m_A = 2 m_{\tilde{ \chi}_1^0}$
which shows that the $A$-funnel solution  is compatible with the QYU condition.
In the   $m_{\tilde{\chi}_1^{\pm}} - m_{\tilde{
\chi}_1^0}$,  $m_{\tilde
{\tau}} - m_{\tilde{ \chi}_1^0}$ planes in Fig.~\ref{plots3}, we draw the unit
slope line which indicates the presence of  stau
coannihilation and bino-Higgsino mixed dark matter scenarios.
We can see how the parameter space is reduced once QYU condition is applied.
In   Fig.~\ref{plots3}  we also present results in the $m_{\tilde{g}} - m_{\tilde{ q}}$  plane. It shows that QYU condition predicts relatively heavy gluinos and the first two family
 squarks ($m_{\tilde{g},\tilde{ q}}> 1$ TeV).

\subsection{Case~~II}

 \begin{figure}[t!]
\centering
\includegraphics[width=9cm]{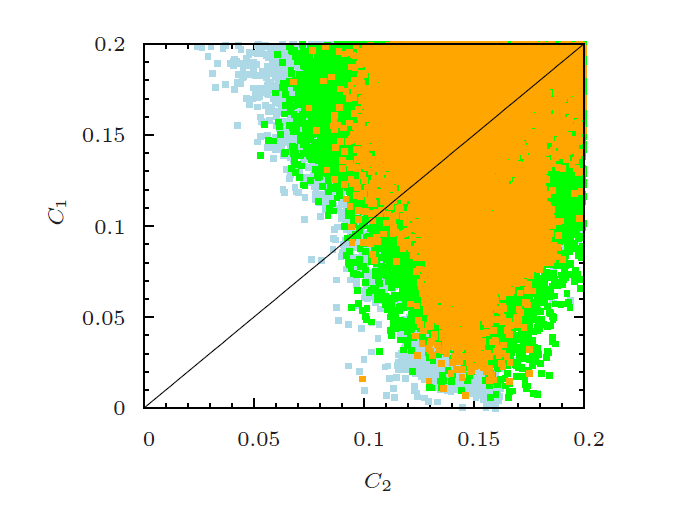}
\caption{
Plots in the $C_1- C_2$  planes for Case II.
Color coding is the same as in Fig. \ref{plots1}.
\label{model2-1}}
\end{figure}

 \begin{figure}[t!]
\centering
\includegraphics[width=15cm]{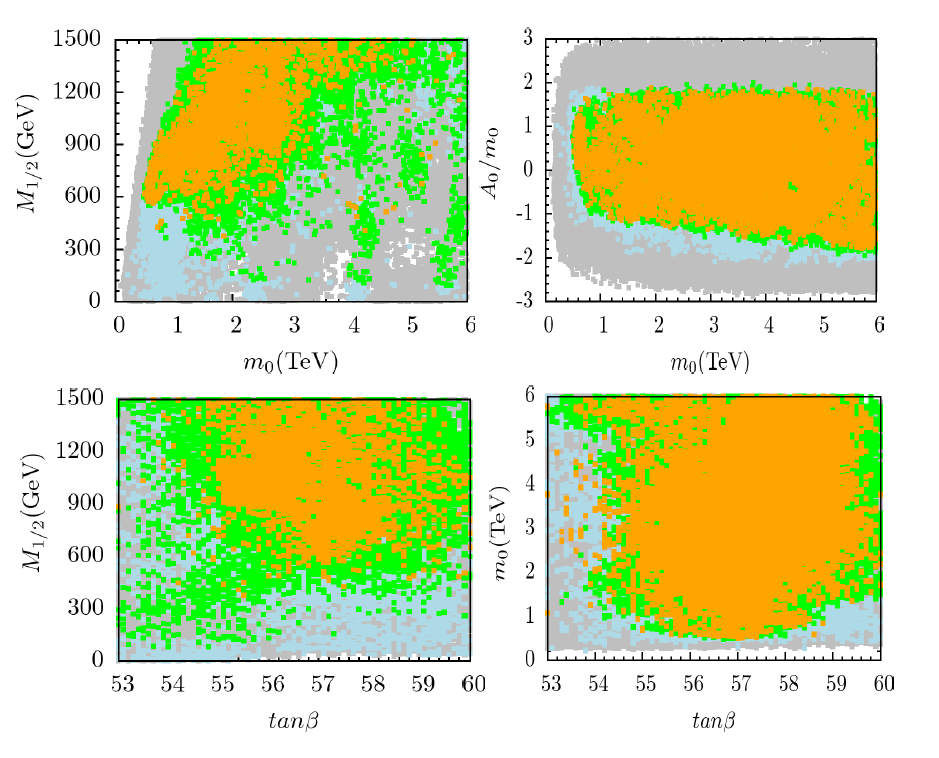}
\caption{
Plots in the $M_{1/2}-m_0$, $M_{1/2}- \tan\beta$,
$A_0/m_0-m_0$, $m_0-\tan\beta$  planes for Case II.
Color coding is the same as in Fig. \ref{plots1}.
\label{model2-2}}
\end{figure}

 \begin{figure}[t!]
\centering
\includegraphics[width=15cm]{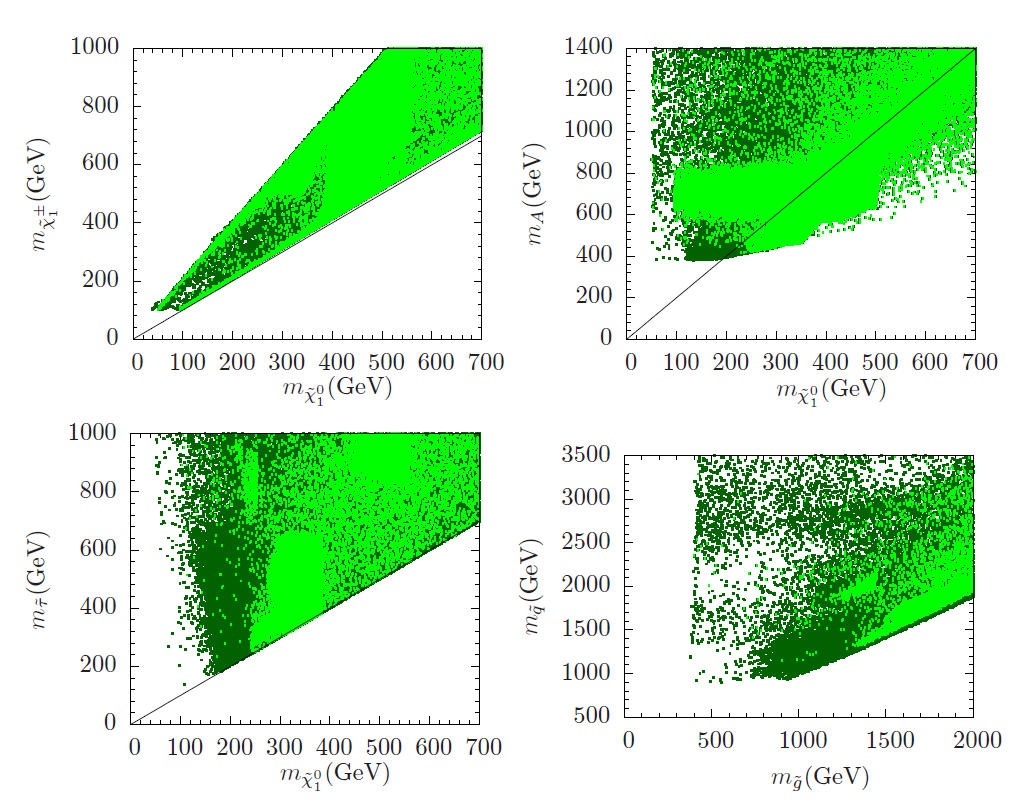}
\caption{
Plots in the $m_{\tilde{\chi}_1^{\pm}} - m_{\tilde{
\chi}_1^0}$, $m_{A} - m_{\tilde{ \chi}_1^0}$,
$m_{\tilde{\tau}} - m_{\tilde{ \chi}_1^0}$ and
$m_{\tilde {q}} - m_{\tilde{ g}}$  planes for Case II.
Color coding is the same as in Fig. \ref{plots3}.
\label{model2-3}}
\end{figure}

As we saw in Fig. \ref{plots1}, the QYU condition
presented in  Eq. (\ref{rela1}) strongly squeezes the allowed
fundamental parameter space (light blue points) compared to the
parameter space for  NUHM1 (gray points) in the absence of this condition. On the other hand,
 the relation presented in Eq. (\ref{rela2})  is quite common in SO(10) model building.
 We find that with completely arbitrary values for the parameters $C_1$ and $C_2$ in Eq. (\ref{rela2}),
   the  allowed parameter space is very similar to what we have for NUHM1.
However, arbitrary values for $C_1$ and $C_2$ contradict our strategy, which
 is to have as small a deviation as possible from exact $t$-$b$-$\tau$ YU.
 Thus, we impose  $C_1<0.2$ and $C_2<0.2$ in case II.

  In Fig. \ref{model2-1} we present our results in the $C_1- C_2$ plane.
 The color coding is the same as in Fig.~\ref{plots1}. The parameter $C_1$ can be as small as 0.01, while it is hard
 to find $ C_{2} $ values less than 0.1.

In Fig.~\ref{model2-2} we present our results in the
$M_{1/2}-m_0$, $M_{1/2}-\tan\beta$, $A_0/m_0-m_0$,
$m_0-\tan\beta$  planes for $C_1<0.2$ and $C_2<0.2$. The color
coding is the same as in Fig.~\ref{plots1}. We can see that the allowed
parameter space is  increased  compared to Case I in
Fig.~\ref{plots1}. The lower bound for  $M_{1/2}$   satisfying
all constraints except the WMAP dark matter relic density bound is almost twice as small
compared to Case I.

The differences in the allowed parameter space between Case I and Case II
become more visible when we compare  Fig. \ref{plots3} and \ref{model2-3}. The LSP neutralino in Case II can be as light as 50
GeV or so, while for Case I the corresponding lower bound is $m_{\chi^{0}_{1}}
\approx 200$ GeV. The $m_{\tilde{\chi}_1^{\pm}} - m_{\tilde{
\chi}_1^0}$ plane shows that there are plenty of bino-Higgsino dark matter solutions.

In Table~\ref{table1} we present some benchmark points for the SO(10)
QYU scenario implemented in NUHM1. All of these points are
consistent with neutralino dark matter and the constraints
mentioned in Section~\ref{constraintsSection}.
For Point 1  bino-Higgsino mixing
plays a major role in giving the correct dark matter relic density. Point 2 corresponds to the $A$-funnel solution,
while point 3 represents the stau coannihilation channel.   Point 4 represents Case II
where $C_1+C_2$ is taken to be minimal, which is equivalent to have minimal deviation between
 $y_t$ and $y_b$ at $M_{\rm G}$.
As expected \cite{Gogoladze:2010ch}, both the spin independent
and spin dependent cross sections of the neutralinos on protons
are larger for Point 1.

\begin{table}[h!]
\centering
\begin{tabular}{lcccc}
\hline
\hline
                 & Point 1 & Point 2     & Point 3  & Point 4  \\
\hline
$m_{0}$          & 3535  & 838.2    & 781.9   & 1123  \\
$M_{1/2} $         & 1585   & 674.30    & 970.40  & 741.5  \\
$\tan\beta$      & 55.1    & 58.1      & 58.2    & 57.5  \\
$A_0/m_0$        & -0.42 & 1.09    & 0.82 & 0.72  \\
$m_{Hu}=m_{Hd}$         & 4149  & 254    & 727   &  1375\\
$m_t$            & 173.1 & 173.1   & 173.1 & 173.1 \\
sgn $\mu$        & +    & +      & +    & +  \\

\hline
$m_h$            & 118.26  & 113.45      & 115.44   & 113.94 \\
$m_H$            & 1714.9  & 546.76     & 703.34   & 476.79\\
$m_A$            & 1703.7  & 543.15     & 698.72   & 473.65\\
$m_{H^{\pm}}$    & 1717.4  & 553.90     & 709.12   & 485.54\\

\hline
$m_{\tilde{\chi}^0_{1,2}}$
                 &623.45, 640.99    &284.27, 542.35  & 414.82, 780.93 & 310.49, 471.73\\
$m_{\tilde{\chi}^0_{3,4}}$
                 &715.69, 1336.9 &960.6, 966.75  &1035.9, 1046.8 &492.56, 628.32 \\

$m_{\tilde{\chi}^{\pm}_{1,2}}$
                 &651.21, 1315.8  &543.08, 967.24  &781.53, 1047.2 &475.95, 620.41 \\
$m_{\tilde{g}}$  & 3522.8     & 1549.9        & 2147.9 & 1699.3 \\

\hline $m_{ \tilde{u}_{L,R}}$
                 &4635.6, 4564.6  & 1613.5, 1571.9  &2086.6, 2016.9 &1875.3, 1833.5 \\
$m_{\tilde{t}_{1,2}}$
                 & 3068.4, 3512.5  & 1292, 1447.6  & 1640.2, 1848.5 &1361, 1534.1 \\
\hline $m_{ \tilde{d}_{L,R}}$
                 &4636.3, 4556.3 & 1615.6, 1567.1    & 2088.2, 2009.6 &1877.1, 1829 \\
$m_{\tilde{b}_{1,2}}$
                 &3488.9, 3605.3 &1377.5, 1458.6    &1759.6, 1843.1 &1483.3, 1531.1 \\
\hline
$m_{\tilde{\nu}_{1}}$
                 & 3678.2       & 946.72         & 1008.6   & 1221.9\\
$m_{\tilde{\nu}_{3}}$
                 &  2821.6      &799.49         &830.01  & 891.31\\
\hline
$m_{ \tilde{e}_{L,R}}$
                &3678.8, 3579.3   &950.80, 874.37  &1013.1, 860.08 & 1224.9, 1155.3\\
$m_{\tilde{\tau}_{1,2}}$
                & 2198, 2820.9  & 533.66, 819.86 & 427.96, 847.75 &530.84, 897.20\\
\hline

$\sigma_{SI}({\rm pb})$
                & $0.17915\times 10^{-7}$ & $0.14364\times 10^{-8}$
                & $0.73529\times 10^{-9}$ & $0.27819\times 10^{-7}$ \\

$\sigma_{SD}({\rm pb})$
                & $0.11476 \times 10^{-4}$ & $0.89996 \times 10^{-7}$
                & $0.88293\times 10^{-7}$ & $0.48096\times 10^{-5}$ \\

$\Omega_{CDM}h^{2}$
                &  0.086      & 0.074     & 0.118 & 0.108 \\

$C$             &  0.15       & 0.21  & 0.19  & $C_1$=0.03,  $C_2$=0.14 \\
\hline
\hline
\end{tabular}
\caption{
Sparticle and Higgs masses (in GeV). All of these benchmark points satisfy the various constraints
mentioned in Section  \ref{constraintsSection} and are compatible with QYU in the NUHM1 framework.
LSP in Point 1  is a bino-Higgsino admixture, while Point 2 corresponds to the $A$-funnel solution.
Point 3 represents the stau coannihilation channel.   Point 4 represents Case II
with minimal value of $C_1+C_2$.
\label{table1}}
\end{table}

%%%%%%%%%%%%%%%%%%%%%%%%%%%%%%%%%%%%%%%%%%%%%%

%%%%%%%%%%%%%%%%%%%%%%%%%%%%%%%%%%%%%%%%%%%

\section{Quasi-Yukawa Unification and Sparticle Spectroscopy: CMSSM \label{ch:cmssm}}

 \begin{figure}[t!]
\centering
\includegraphics[width=15cm]{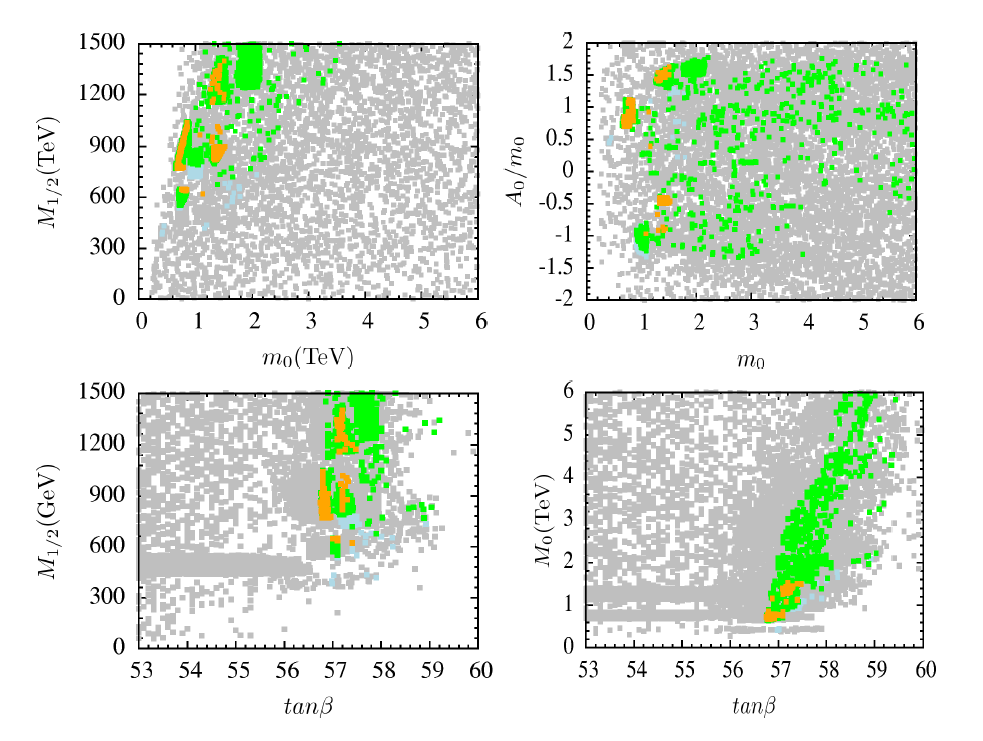}
\caption{
Plots in the $M_{1/2}-m_0$, $M_{1/2}- \tan\beta$,
$A_0/m_0-m_0$, $m_0-\tan\beta$  planes for the CMSSM.
Color coding is the same as in Fig. \ref{plots1}.
\label{cmssm-1}}
\end{figure}

\begin{figure}[t!]
\centering
\includegraphics[width=15cm]{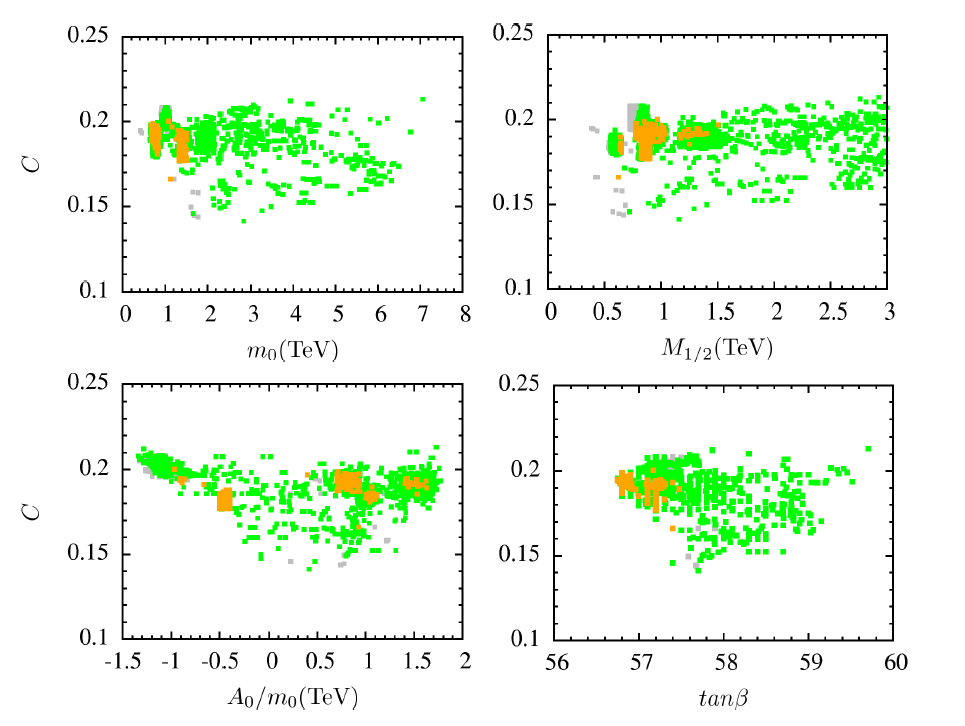}
\caption{
Plots in the $C-m_0$, $C-M_{1/2}$, $C-A_0/m_0$, $C-\tan\beta$  planes for the CMSSM.
Color coding is the same as in Fig. \ref{plots1}.
\label{cmssm-2}}
\end{figure}

 \begin{figure}[t!]
\centering
\includegraphics[width=15cm]{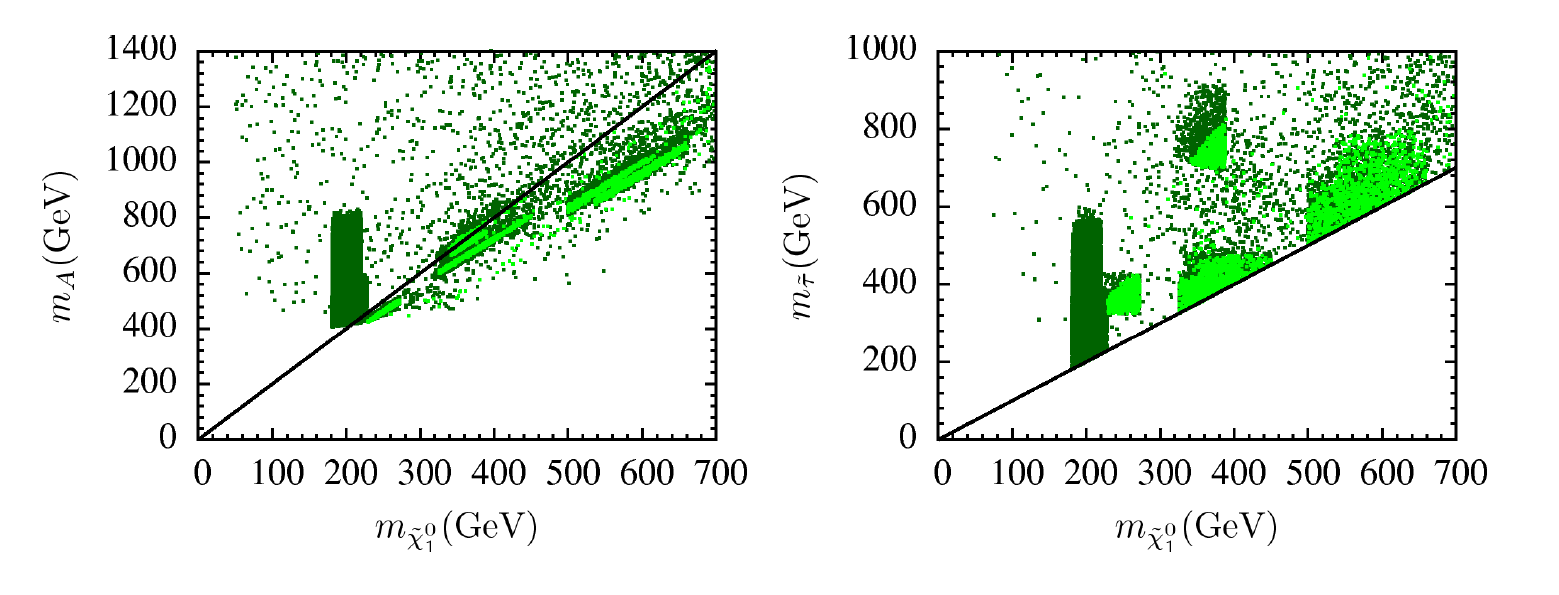}
\caption{
Plots in the $m_{A} - m_{\tilde{ \chi}_1^0}$ and
$m_{\tilde{\tau}} - m_{\tilde{ \chi}_1^0}$ and
  planes for the CMSSM case.
Color coding is the same as in Fig. \ref{plots3}.
\label{cmssm3}}
\end{figure}

\begin{table}[h!]
\centering
\begin{tabular}{lcc}
\hline
\hline
                 & Point 1 & Point 2\\
\hline
$m_{0}$          & 692.6  & 1395   \\
$M_{1/2} $         & 804.3   & 855.1 \\
$\tan\beta$      & 56.8    & 57.2   \\
$A_0/m_0$        & 0.76 & -0.42     \\
$m_t$            & 173.1 & 173.3   \\
sgn $\mu$        & +    & +      \\

\hline
$m_h$            & 117.5  & 119  \\
$m_H$            & 635.1  & 715.3 \\
$m_A$            & 630.9  & 710.6\\
$m_{H^{\pm}}$    & 641.5  & 721.1\\

\hline
$m_{\tilde{\chi}^0_{1,2}}$
                 &340.5, 641.5    &368.2, 698.3  \\
$m_{\tilde{\chi}^0_{3,4}}$
                 &866.1, 878.9 &1025.3, 1034   \\

$m_{\tilde{\chi}^{\pm}_{1,2}}$
                 &642, 879.1  &699, 1034.2  \\
$m_{\tilde{g}}$  & 1806.4     & 1952.1        \\

\hline $m_{ \tilde{u}_{L,R}}$
                 &1768.8, 1712.6  & 2210, 2162.9  \\
$m_{\tilde{t}_{1,2}}$
                 & 1362.8, 1561.8  & 1544.4, 1800.9   \\
\hline $m_{ \tilde{d}_{L,R}}$
                 & 1770.8, 1706.9  & 2211.5, 2157.5     \\
$m_{\tilde{b}_{1,2}}$
                 & 1486, 1560.1 &1746.6, 1827.2     \\
\hline
$m_{\tilde{\nu}_{1}}$
                 & 870.8       & 1501.7         \\
$m_{\tilde{\nu}_{3}}$
                 &  742.7      &1239.3         \\
\hline
$m_{ \tilde{e}_{L,R}}$
                &875.6, 754   &1504.2, 1429.4 \\
$m_{\tilde{\tau}_{1,2}}$
                & 358.5, 760  & 755.1, 1245.3 \\
\hline

$\sigma_{SI}({\rm pb})$
                & $0.13\times 10^{-8}$ & $0.56\times 10^{-9}$ \\

$\sigma_{SD}({\rm pb})$
                & $0.27\times 10^{-6}$ & $0.13 \times 10^{-6}$\\

$\Omega_{CDM}h^{2}$
                &  0.11      & 0.10  \\

$C$             &  0.19       & 0.18   \\
\hline
\hline
\end{tabular}
\caption{
Sparticle and Higgs masses (in GeV). These benchmark points satisfy the various constraints
mentioned in Section  \ref{constraintsSection} and are compatible with QYU in the CMSSM framework.
 Point 1 corresponds to the stau coannihilation channel, while
Point 2 represents the $A$-funnel solution.
\label{table2}}
\end{table}
%%%%%%%%%%%%%%%%%%%%%%%%%%%%%%%%%%%%%%%%%

 The CMSSM is one of the most popular frameworks for  studying the  low scale sparticle spectroscopy, and we
  therefore, discuss it here in light of  QYU. Following the NUHM1 case,  we perform random scans for the parameter range presented in Eq. (\ref {parameterRange})
with the additional constraint $m_0=m_{H_{u}}=m_{H_{d}}$.

In Fig.~\ref {cmssm-1} we present our results in the
$M_{1/2}-m_0$, $M_{1/2}-\tan\beta$, $A_0/m_0-m_0$ and
$m_0-\tan\beta$  planes for the CMSSM case. The color
coding is the same as in Fig.~\ref{plots1}. We can see that the allowed
parameter space is  restricted   compared to what we have for Case I in
Fig.~\ref{plots1}. In good approximation we can say that QYU predicts $\tan\beta \approx 57$
which is compatible with all collider and WMAP bounds.  Also  QYU prefers smaller value for $m_0< 2$ TeV while precise  $t$-$b$-$\tau$ YU with universal gaugino masses prefers $m_{0}\gtrsim 6$ TeV  \cite{Baer:2008jn}. There is no preference for $A_0/m_0$ value in contrast to the precise  $t$-$b$-$\tau$ YU.

In Fig. \ref{cmssm-2}, we present our results in the $C-m_0$, $C-M_{1/2}$, $C-A_0/m_0$, $C-\tan\beta$  planes.
Color coding is the same as Fig.~\ref{plots1}.
A comparison with Fig. \ref{plots2} shows that the allowed value of $ C $ is significantly constrained to $ \sim 0.2 $.
 In contrast to NUHM1, we are unable to find bino-Higgsino dark matter in the CMSSM framework with QYU.
But as we can see in Fig. \ref{cmssm3}, there are plenty of  stau coannihilation and $A$-funnel solution.

In Table~\ref{table2} we present some benchmark points for
QYU in the CMSSM framework. All of these points are
consistent with neutralino dark matter and the constraints
mentioned in Section~\ref{constraintsSection}.
 Point 2 corresponds to the stau coannihilation channel and Point 3 represents  $A$-funnel solution.

\section{Quasi-Yukawa Unification  and Dark Matter Detection \label{ch:dark}}

\begin{figure}[t!]
\centering
\includegraphics[width=16cm]{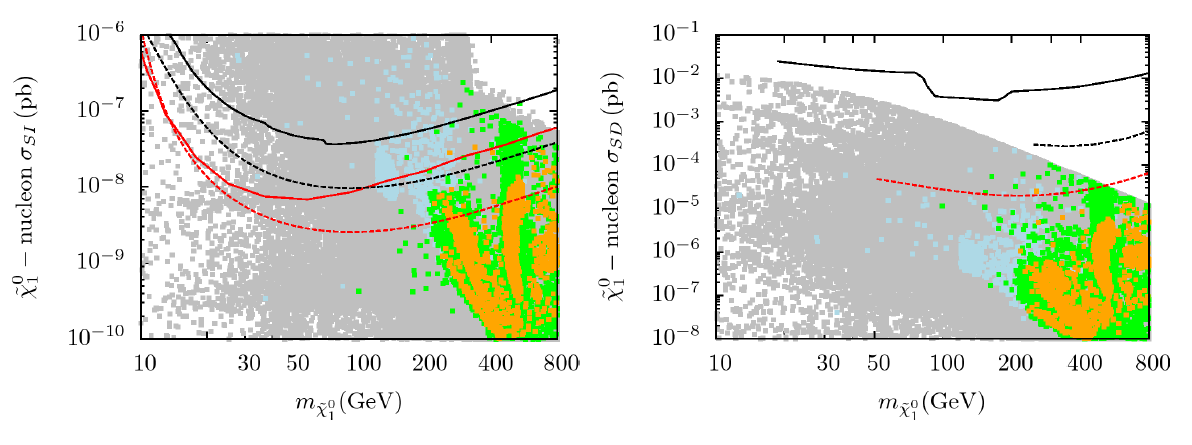}
\caption{
Plots in the $\sigma_{\rm SI}$ -$m_{\tilde{\chi}_1^{0}}$ and $\sigma_{\rm SD}$ -
$m_{\tilde{\chi}_1^{0}}$
planes for Case I. Color coding is the same as in Fig.~\ref{plots1}.
 In the $\sigma_{\rm SI}$ -
$m_{\tilde{\chi}_1^{0}}$ plane, the current and future bounds are represented by black (solid and dashed) lines for the CDMS experiment and by red (solid and dashed) lines for the Xenon experiment. In the $\sigma_{\rm SD}$ -
$m_{\tilde{\chi}_1^{0}}$ plane we show the current bounds from Super
K (black line) and IceCube (dotted red line) and future reach of
IceCuce DeepCore (red solid line).
\label{plots4}}
\end{figure}

\begin{figure}[t!]
\centering
\includegraphics[width=16cm]{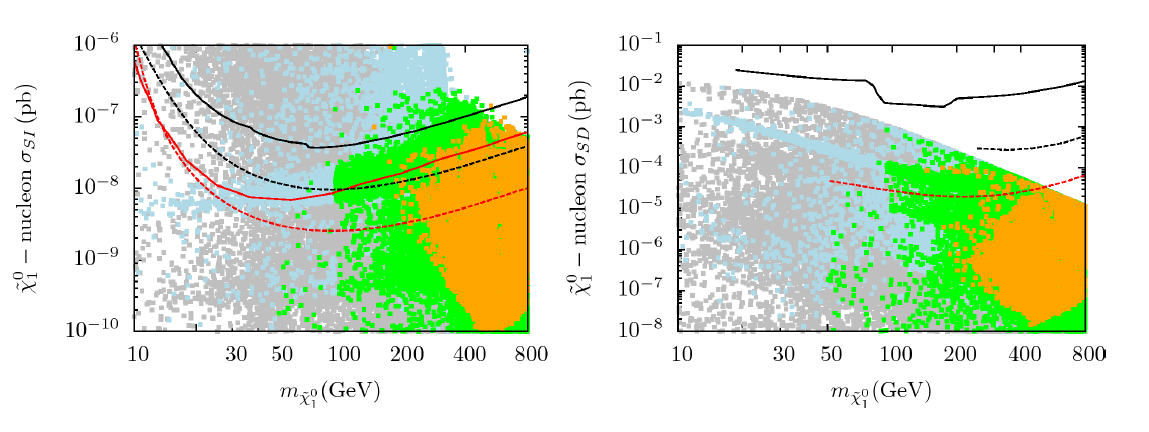}
\caption{
Plots in the $\sigma_{\rm SI}$ -$m_{\tilde{\chi}_1^{0}}$ and $\sigma_{\rm SD}$ -
$m_{\tilde{\chi}_1^{0}}$
planes for Case II. Color coding is the same as in Fig.~\ref{plots1}
 In the $\sigma_{\rm SI}$ -
$m_{\tilde{\chi}_1^{0}}$ plane, the current and future bounds are represented by black (solid and dashed) lines for the CDMS experiment and by red (solid and dashed) lines for the Xenon experiment. In the $\sigma_{\rm SD}$ -
$m_{\tilde{\chi}_1^{0}}$ plane we show the current bounds from Super
K (black line) and IceCube (dotted red line) and future reach of
IceCuce DeepCore (red solid line). \label{model2-4}}
\end{figure}

In light of the recent results by the CDMS-II \cite{Ahmed:2009zw}
and Xenon100 \cite{Aprile:2010um}  experiments, it is important to
see if  QYU, within the NUHM1 and CMSSM frameworks presented in
this paper, is testable from the perspective of direct and indirect
detection experiments. The question of interest is whether $\mu \sim
M_1$ is consistent with QYU, as this is the
requirement to get a bino-Higgsino admixture for the lightest
neutralino which, in turn, enhances both the spin dependent and spin
independent neutralino-nucleon scattering cross
sections~\cite{Gogoladze:2010ch}. In Fig.\ref{plots4} for Case I
and Fig.\ref{model2-4} for Case 2, we show the spin independent
and spin dependent cross sections as a function of the LSP neutralino
mass. In the $\sigma_{\rm SI}$ -
$m_{\tilde{\chi}_1^{0}}$ plane, the current and future bounds are represented by black (solid and dashed) lines for the CDMS experiment \cite{Bruch:2010eq} and by red (solid and dashed) lines for the Xenon experiment. The color coding is the same as in
Fig.~\ref{plots1}. For Case II a small region of the parameter
space consistent with QYU  and the experimental
constraints discussed in Section~\ref{constraintsSection}  is at the
exclusion limits set by the recent CDMS and XENON experiments.
Thus, the ongoing and planned direct detection experiments will play
a vital role in testing QYU. Note that the above remarks only apply to NUHM1 which contains bino-Higgsino dark matter solutions.

For the spin dependent cross section, we show in
Figs.\ref{plots4} and \ref{model2-4} the current bounds from the
Super-K (black line) \cite{Desai:2004pq} and IceCube (dotted red line) \cite{Abbasi:2009uz}
experiments and the projected future reach of IceCube DeepCore (red solid line). The current
Super-K and IceCube bounds are not stringent enough to rule out the
parameter space characteristic for NUHM1
 with QYU. However,
from Figs.\ref{plots4}  and \ref{model2-4} we observe that the
IceCube DeepCore experiment should be able to constrain some region
of the parameter space, especially for Case II.

%%%%%%%%%%%%%%%%%%%%%%%%%%%%%%%%%%%%%%%%%%%%%%%%%

%

%%%%%%%%%%%%%%%%%%%%%%%%%%%%%%%%%%%%%%%%%%%%%%%%%%

\begin{figure}[t!]
\centering
\includegraphics[width=15.5cm]{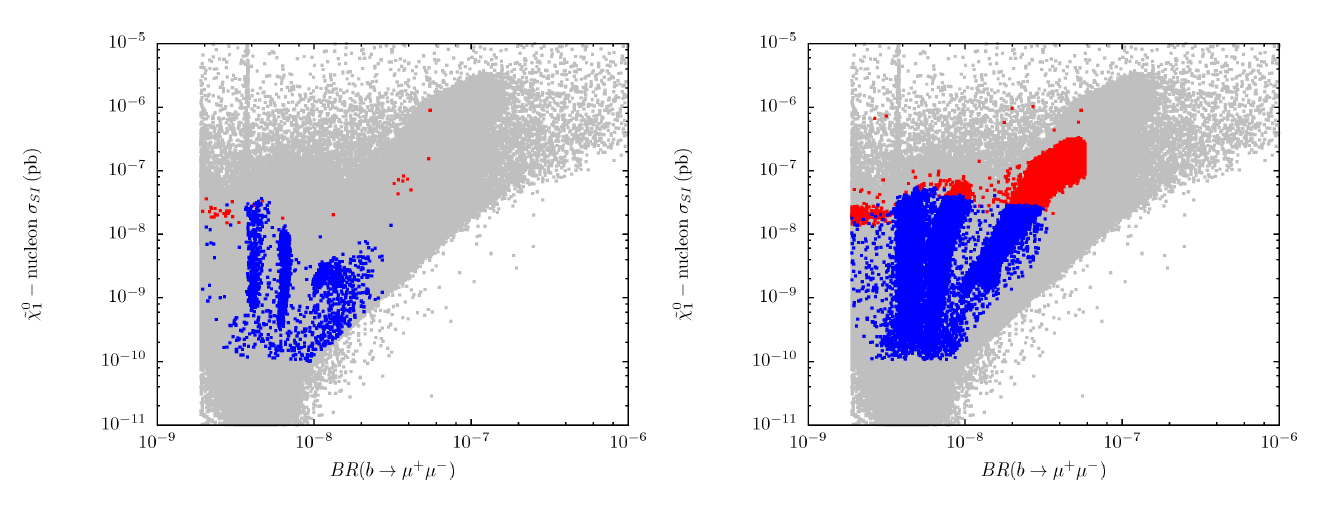}
\caption{ Plots in  $\sigma_{\rm SI}-BR(B_s\rightarrow \mu^+ \mu^-)$  for case I (left) and Case II (right) panels in NUHM1 parameter space.
Gray points are consistent with REWSB and neutralino LSP.
Red points are subset of gray points which  are excluded by Xenon100 and CDMS-II experiments.
Blue points belong to a subset of gray points consistent with QYU and the experimental constraints discussed  in Section 3. Blue points can  be tested by Xenon1T experiment \cite{XENON1T}.
\label{fig-1}}
\end{figure}

In Figure \ref{fig-1}, we show the spin independent cross section as a function of  $BR(B_s\rightarrow \mu^+ \mu^-)$   for case I (left panel) and Case II (right panel)
in the NUHM1 parameter space. Gray points are consistent with REWSB and neutralino LSP.
Red points are a subset of gray points which  are excluded by Xenon100 and CDMS-II experiments.
Blue points belong to a subset of gray points and they can be tested by Xenon100 experiment. Red and blue points are both consistent with QYU and the experimental constraints discussed  in Section 3. The blue points correspond to the orange points which are located between the dashed and solid red lines in Figure 10 and Figure 11 in  $\sigma_{\rm SI}$ -$m_{\tilde{\chi}_1^{0}}$   panel.  According to LHCb results, the lower bound on the rare decay $BR(B_s\rightarrow \mu^+ \mu^-)$ approaches its SM limit, which means that supersymmetric contribution has to get smaller. This is the reason why we have a big interval for the neutralino nucleon cross section when
$2.3\times10^{-9}< BR(B_s\rightarrow \mu^+ \mu^-)< 1.2\times10^{-8}$.

%%%%%%%%%%%%%%%%%%%%%%%%%%%%%%%%%%%%%%%%%%%%%%%%%%%%
%%%%%%%%%%%%%%%%%%%%%%%%%%%%%%%%%%%%%%%

%%%%%%%%%%%%%%%%%%%%%%%%%%%%%%%%%%%%%
\section{Conclusion
\label{ch:conclusions}}

We have explored quasi or approximate third family $t$-$b$-$\tau$ YU predicted in a class of realistic SO(10) and
$SU(4)_c \times SU(2)_L \times SU(2)_R$  models. We find that QYU when implemented in a NUHM1 setup is compatible with the bino-Higgsino dark matter as well as the stau coannihilation and the $A$-funnel solutions. We could not identify a bino-Higgsino dark matter solution in the CMSSM case in the parameter range that we have examined. The MSSM parameter tan $\beta$, as expected, turns out to be fairly large in QYU models, of order 54-60. The prospects for testing these ideas in the ongoing experiments is briefly discussed.

%%%%%%%%%%%%%%%%%%%%%%%%%%%%%%%
{\bf Note added}: A similar analysis in the CMSSM framework is being carried out by N. Karagiannakis, G. Lazarides and C. Pallis (private communication).

%%%%%%%%%%%%%%%%%%%%%%%%%%%%%%%%%%
\section*{Acknowledgments}
We thank Rizwan Khalid, C. Pallis and Shabbar Raza for valuable discussions. This work
is supported in part by the DOE Grant No. DE-FG02-91ER40626
(I.G., C.U., and Q.S.) and GNSF Grant No. 07\_462\_4-270 (I.G.).

%%%%%%%%%%%%%%%%%%%%%%%%%%%%%%%%%

\end{document}